\documentclass[aps,pre,reprint,onecolumn,notitlepage]{revtex4-1}
\usepackage{epsfig,graphics,amssymb,amsmath,subeqnarray,color,bm}
\usepackage[toc,page]{appendix}
\usepackage{mathtools}
\usepackage{xcolor}
\usepackage{float}

\numberwithin{equation}{section}
\usepackage{graphicx}

\def \lb{\left}
\def \rb{\right}

\def \Deltab{\bar{\Delta}}

\def \bF{\mathbf{F}}

\def \bG{\mathbf{G}}

\def \bI{\mathbf{I}}

\def \bR{\mathbf{R}}

\def \bU{\mathbf{U}}

\def \bFh{\hat{\mathbf{F}}}

\def \bUh{\hat{\mathbf{U}}}

\def \be{\mathbf{e}}

\def \bu{\mathbf{u}}

\def \bx{\mathbf{x}}

\def \bzero{\mathbf{0}}

\def \Ff{\mathcal{F}}

\def \eq{\text{eq}}

\begin{document}
\title{Hydrodynamic interactions of cilia on a spherical body}
\author{Babak Nasouri}
\affiliation{
Department of Mechanical Engineering,  
University of British Columbia,
Vancouver, B.C., V6T 1Z4, Canada}
\author{Gwynn J. Elfring}\email{Electronic mail: gelfring@mech.ubc.ca}
\affiliation{
Department of Mechanical Engineering,  
University of British Columbia,
Vancouver, B.C., V6T 1Z4, Canada}

\begin{abstract} 
Microorganisms develop coordinated beating patterns on surfaces lined with cilia known as metachronal waves. For a chain of cilia attached to a flat ciliate, it has been shown that hydrodynamic interactions alone can lead the system to synchronize. However, several microorganisms possess a curve shaped ciliate body and so to understand the effect of this geometry on the formation of metachronal waves, we evaluate the hydrodynamic interactions of cilia near a large spherical body. Using a minimal model, we show that for a chain of cilia around the sphere, the natural periodicity in the geometry leads the system to synchronize. We also report an emergent wave-like behavior when an asymmetry is introduced to the system.
\end{abstract}
\maketitle
\section{Introduction}
To propel themselves in a low-Reynolds-number regime \cite{purcell1977}, many microorganisms use small whip-like extensions, called flagella (when they possess one or two) or cilia (when they possess many) \cite{lodish2000,brennen1977}. The motion of cilia is controlled by  ATP-fuelled motor proteins which exert a driving force on the cilia by converting chemical energy in the cell \cite{roberts2013}. The cyclic motion of each cilium in a chain can form a coordinated pattern of beating, wherein each pair of neighboring cilia are orbiting with a constant, non-zero, phase difference \cite{knight1954}. As a result of this synchrony, the tips of cilia form a moving wave, known as a metachronal wave \cite{gueron1997}. By forming metachronal waves, the microorganism minimizes the required energy for beating \cite{gueron1999}, which enhances the efficiency of the motion \cite{kim2006}. In addition to providing a means of locomotion, cilia in the human body filter air flow channels in the lung from the harmful inhaled material \cite{shah2009}, and also play a crucial role in breaking the left-right symmetry in human embryonic development \cite{hirokawa2009}.

Several analytical studies have shown that hydrodynamic interactions alone can lead to synchronization (zero phase difference) or phase-locking (constant non-zero phase difference) for model systems of two flagella \cite{vilfan2006,elfring2011b,uchida2011} or many cilia \cite{lagomarsino2003,niedermayer2008}. Using minimal models, it was observed that certain conditions, be they elastic deformations of trajectory \cite{niedermayer2008}, or shape \cite{elfring2010}, or a certain forcing profile \cite{uchida2011}, may be required to reach such phase-locking or synchronization. Recent experimental studies have also confirmed the hydrodynamic synchronization of micro-scale oscillators in natural systems. Using high-speed video microscopy, it was shown that beating flagella on \textit{Chlamydomonas reinhardtii} \cite{goldstein2009,polin2009} and \textit{Volvox carteri} \cite{brumley2012,brumley2015}, exhibit a synchronization due to hydrodynamic interactions. Similar synchrony was observed in model colloidal systems where two spheres were oscillating on linear \cite{kotar2010,bruot2011} or circular \cite{box2015} trajectories and each sphere was driven by optical tweezers.

In a ciliary array, the distribution of cilia as well as the details of the ciliate body affect the behavior of the dynamical system \cite{golestanian2011,elgeti2015,bruot2015}. Niedermayer \textit{et al.} reported that introducing radial flexibility to the circular trajectory of two orbiting beads leads to synchronization, but that a non-periodic array of such beads cannot reach stable collective phase-locking \cite{niedermayer2008}. They also showed that marginally-stable metachronal waves are formed only when the cilia are distributed in a periodic fashion. More recently it was observed that an open-ended array of cilia can indeed form robust metachronal waves if the cilia beat perpendicular to the ciliate boundary \cite{brumley2012}. It has also been shown that the presence of a large body near an array of linearly oscillating beads is necessary for emergence of metachronal waves \cite{wollin2011}. The bounding surface restricts the range of hydrodynamic interactions of the beads and leads the system to a collectively phase-locked state. The emerging picture from the literature is that the stability and existence of metachronal waves depends on the geometry of the cilia and ciliate body. Notably however, in many ciliates in nature the cilia are continuously distributed about a closed curved body such as on \textit{paramecia} or \textit{volvox}, and this imposes a natural periodicity to the dynamical system and mediates the hydrodynamic interactions of the ciliary chains in a way that is yet to be understood.  

In this paper we investigate the effects of a large curved ciliate body on the hydrodynamic interactions of cilia in a viscous fluid. Following the work of Niedermayer \textit{et al.} who studied interactions of cilia above a flat wall \cite{niedermayer2008}, we use the discrete-cilia model \cite{blake1971,liron1976} where each cilium is replaced by a single sphere and assume that a constant tangential forcing is applied by the dynein motors. We first present an analysis of the interactions of two cilia and then build up our model of a chain of cilia around a large spherical body. We show that the radial flexibility in the trajectories can lead the system to synchronize similar to the case of cilia near a flat boundary \cite{niedermayer2008}. Furthermore, we show that with this model, the only stable fixed point for a chain of identical cilia is when all cilia are in phase (synchronized). Finally, we demonstrate an emergent wave-like behavior of the cilia in response to an imposed asymmetry in the beating rate of one cilium.
\section{Motion of a single cilium}
\begin{figure}
\begin{center}
\includegraphics[scale=0.7]{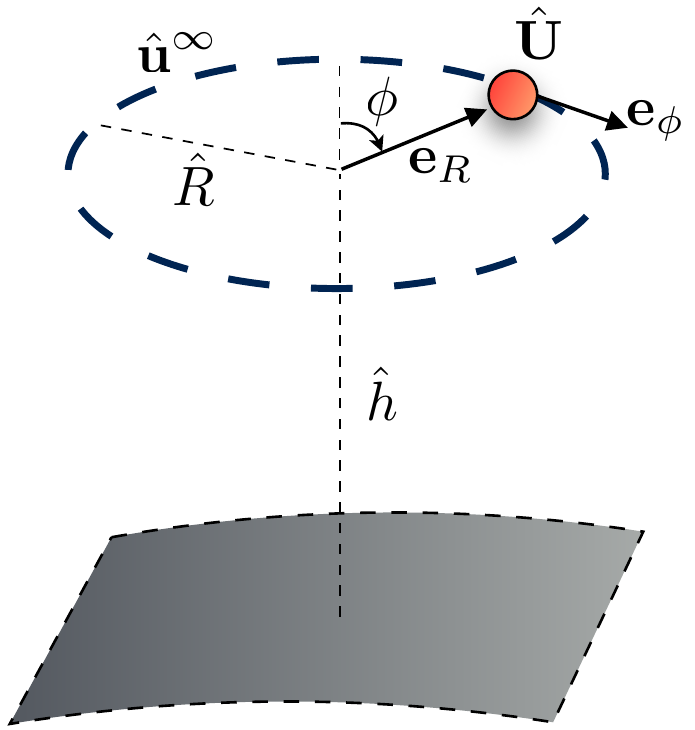}
\caption{A schematic of the motion of a model cilium near a spherical body. The circular trajectory has a radius of $\hat{R}$ and its center is at distance $\hat{h}$ from the boundary. The cilium moves with velocity $\hat{\bU}$ through a fluid with velocity $\hat{\bu}^{\infty}$. In this study, $\phi$ indicate the instantaneous phase of the cilium and the vectors $\be_{\phi}$ and $\be_{R}$ show the tangential and radial directions of the motion, respectively.}
\label{cilium}
\end{center}
\end{figure}

We model the cilium as a single sphere of radius $\hat{a}$ undergoing a circular orbit, of radius $\hat{R}$, whose center is at distance $\hat{h}$ from the ciliate body as shown in Fig.~\ref{cilium}. Dynein motors drive the motion of the cilium and, in a viscous fluid at small scales, this forcing is balanced entirely by the hydrodynamic drag,
\begin{align}
\label{balance}
\hat{\bF}^{m}+\hat{\bF}^{D}=\bzero.
\end{align}
A simple model of the forcing stipulates a constant tangential driving force \cite{lenz2006}, $\hat{F}^{\text{dr}}$, and an elastic restoring force that keeps the cilia moving along a preferred path (of radius $\hat{R}_0$) \cite{niedermayer2008}, such that
\begin{align}
\label{motion1}
\hat{\bF}^{m}=\hat{F}^{\text{dr}}\mathbf{e}_{\phi}-\hat{k}\lb(\hat{R}-\hat{R}_{0}\rb)\mathbf{e}_{R},
\end{align}
where $\hat{k}$ is the stiffness of the cilia. The drag force, $\bFh^D$, for the rigid body translation of a sphere at velocity $\bUh$ is given by the drag law
\begin{align}
\label{motion2}
\hat{\bF}^{D}=-\hat{\bR}_{FU}\cdot\lb(\hat{\bU}-\Ff\lb[\hat{\bu}^{\infty}\rb]\rb),
\end{align}
where $\bu^\infty$ is the background flow and the Fax\'en operator is $\Ff = 1+\frac{\hat{a}^2}{6}\nabla^2$ \cite{batchelor1972}. The resistance tensor for a sphere moving parallel to a wall is $\hat{\bR}_{FU}=6\pi\hat{\mu} \hat{a}\lb(\bI + {O}(\hat{a}^3/\hat{h}^3)\rb)$ \cite{happel1981}. In our analysis, we assume the thickness of a cilium is much smaller than its length so that in our minimal model $\hat{a}\ll\hat{h}$, therefore the effect of wall on the hydrodynamic resistance shall be neglected.

As a starting point, we examine the motion of a single cilium in the absence of other cilia. The background flow field is then zero and the cilium orbits strictly on its preferred circular path. In this case, equation \eqref{motion1} leads to a steady state solution
\begin{align}
\dot{\hat{\phi}}_{ss}&=\dfrac{\hat{F}^{\text{dr}}}{6\pi\hat{\mu} \hat{a}\hat{R}_0}\equiv\hat{\omega},\\
\dot{\hat{R}}_{ss}&=0,\quad \hat{R}_{ss}=\hat{R}_{0},
\end{align}
where the over-dot indicates differentiation with respect to time and $\hat{\omega}$ defined as the intrinsic angular velocity of the cilium. Using the reported values for the bending rigidity of a cilium \cite{okada1999,camalet2000}, Niedermayer \textit{et al.} \cite{niedermayer2008} noted that radial relaxation is much faster than a period of rotation, namely $\hat{F}^{dr}/\hat{k} \hat{R}_0\ll 1$, and so a quasi-static assumption may be employed for the radial dynamics of the system.
\section{Interactions of two cilia}
\begin{figure}
\begin{center}
\includegraphics[scale=0.6]{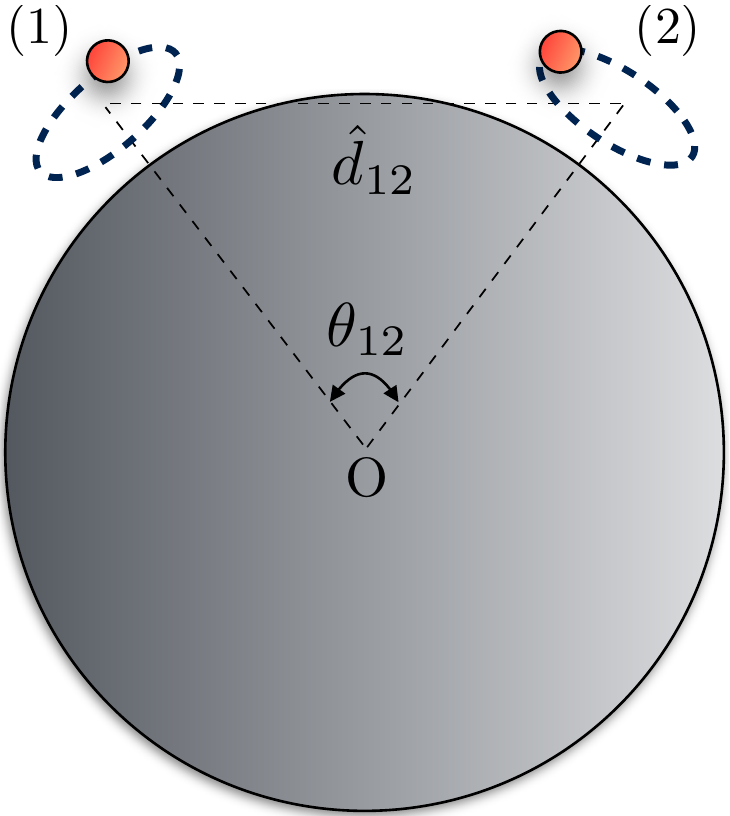}
\caption{A system of two cilia around a spherical body of radius $\hat{A}$. In this figure, $\hat{d}_{12}$ is the distance and $\theta_{12}$ is the angle between center of the trajectories.}
\label{single-cilium}
\end{center}
\end{figure}
We now consider a system of two cilia around a spherical body, with trajectories centered at a distance $\hat{d}_{12}$ as shown in Fig.~\ref{single-cilium}. The velocity of each cilium can be written as $\hat{\bU}_{i}=\hat{R}_{i}\dot{\hat{\phi}}_{i}\be_{\phi_{i}}+ \dot{\hat{R}}_{i}\be_{R_{i}}$, where $i\in\{1,2\}$. The motion of each cilium in this case is affected by the background flow field induced by the other cilium. The ciliate body is considerably larger than the thickness of a cilium, $\hat{a}\ll \hat{A}$, where $\hat{A}$ is the radius of the spherical body. We also assume the cilia are far apart from one another ($\hat{d}_{12}\gg\hat{h}, \hat{R_0}$) \cite{okada1999}, so that far-field approximations for the induced flow fields may be employed. Under these assumptions one can model the flow field due to the motion of a sphere by a point force (or Stokeslet) to leading order while the no-slip boundary condition on the surface of the spherical body is satisfied by an image Stokeslet set in the body. The background flow field on cilium (1), induced by cilium (2) is 
\begin{align}
\label{field}
\hat{\bu}^\infty(\hat{\bx}_1)=\dfrac{1}{8\pi\hat{\mu}}\lb(\hat{\bG}(\hat{\bx}_1,\hat{\bx}_{2})+\hat{\bG}^{*}(\hat{\bx}_1,\hat{\bx}^{*}_{2})\rb)\cdot\hat{\bF}^{m}_{2},
\end{align}
where $\hat{\bF}^{m}_{2}$ refers to the driving force of the cilium (2), $\hat{\bx}_1$ and $\hat{\bx}_2$ indicate the location of each cilium, $\hat{\bG}$ is the Oseen tensor and $\hat{\bG}^{*}$ is the Blake's solution for the image Stokeslet \cite{happel1981,blake1971}, at a point $\hat{\bx}^{*}_{2}=(\hat{A}^2/|\hat{\bx}_{2}|^2)\hat{\bx}_{2}$ located to satisfy the no-slip condition at the spherical boundary \cite{spagnolie2015}.

Before going further, we non-dimensionalize all equations by scaling lengths by the radius of the spherical body, $\hat{A}$, and rates by the average angular velocity of the two cilia, $\bar{\hat{\omega}}$. We drop the notation $(~\hat{}~)$ for the dimensionless quantities defined by $\hat{a}=\hat{A}a$, $\hat{R}_{j}=\hat{A}R_{j}$, $\hat{h}=\hat{A}h$, $\hat{d}_{12}=\hat{A}d_{12}$, $\hat{\omega}_{j}=\bar{\hat{\omega}}\omega_{j}$ and $\dot{\hat{\phi}}_{j}=\bar{\hat{\omega}}\dot{\phi}_{j}$, $\hat{\bU}_{j}=(\hat{A}\bar{\hat{\omega}})\bU_{j}$, $\dot{\hat{R}}_{j}=(\hat{A}\bar{\hat{\omega}})\dot{R}_{j}$, $\hat{t}=(1/\bar{\hat{\omega}})t$ and $\hat{\bR}_{FU}=(\hat{k}/\bar{\hat{\omega}}){\bR}_{FU}$. A dimensionless parameter $\kappa=\hat{k}/6\pi\hat{\mu} \hat{a}\bar{\hat{\omega}}\gg 1$ which indicates the ratio of the elastic restoring force to the hydrodynamic drag force then naturally arises. We assume the dimensionless length scales are ordered as follows, $a \ll \{h,R_0\}\ll 1$ and take $R_0=O(h)$ \cite{okada1999}. Since the radius of the trajectory is small compared to the scale of the body, we may write $d_{12}=2\sin(|\theta_{12}|/2)+O(h)$.

\begin{figure}
\begin{center}
\includegraphics[scale=0.4]{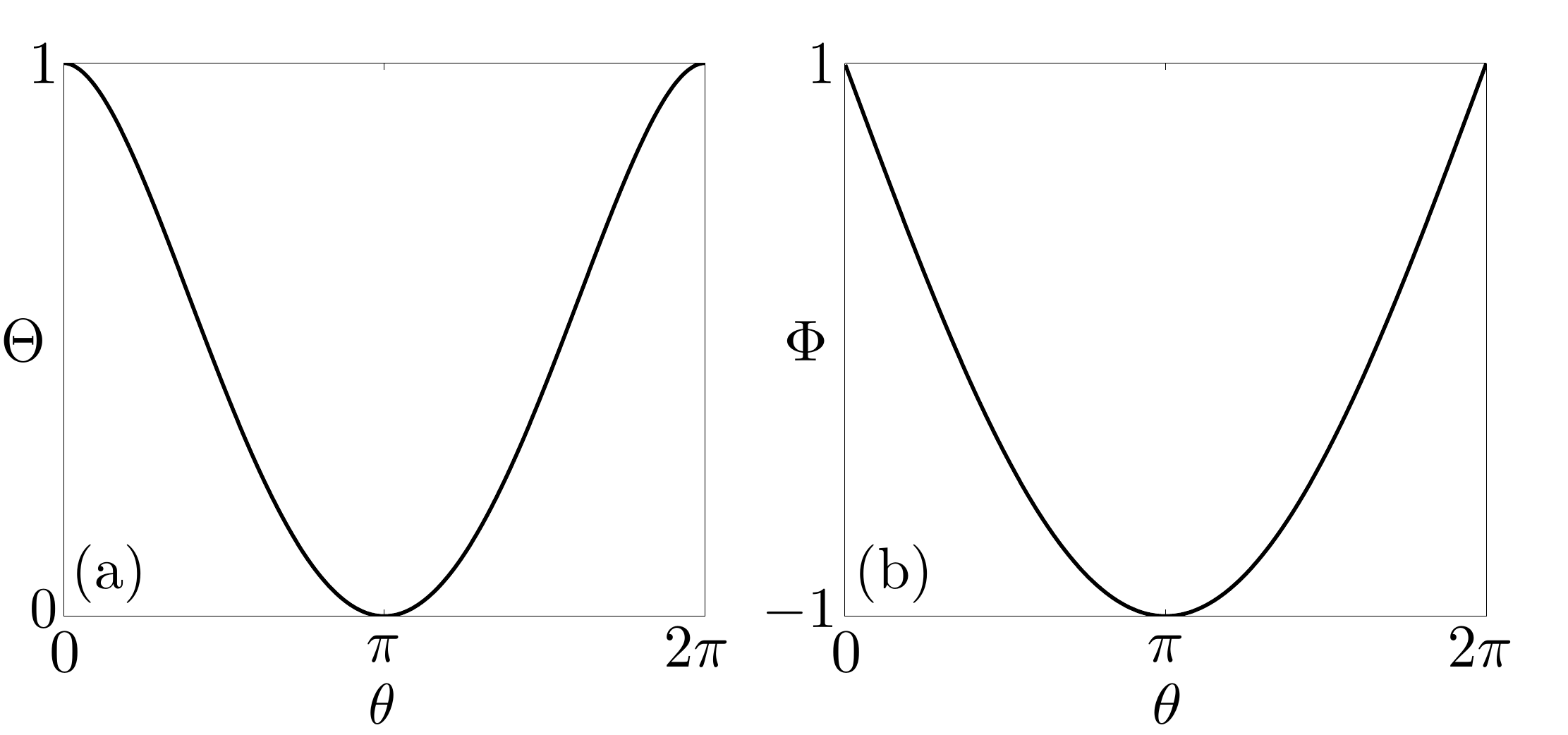}
\caption{Geometric terms (a) $\Theta$ and (b) $\Phi$ as functions of the angle between cilia.}
\label{Theta}
\end{center}
\end{figure}

In these limits, by using the description for background flow field in equation \eqref{field}, the motion equation \eqref{motion1} can be solved for case of two neighboring cilia, asymptotically. The evolution equations are then to leading order
\begin{align}
\label{motion-1}
\dot{\phi}_{1}&= \omega_{1} + \rho\omega_2 S_{12}- \dfrac{\rho\omega_1\omega_2}{\kappa} L_{12},\\
\label{motion-2}
\dot{\phi}_{2}&= \omega_2 +\rho\omega_1 S_{21} - \dfrac{\rho\omega_1\omega_2}{\kappa} L_{21},\\
R_{1}&=R_0+\dfrac{\rho R_0 \omega_2}{\kappa}L_{12},\\
R_{2}&=R_0+\dfrac{\rho R_0 \omega_1}{\kappa}L_{21},
\end{align}
where $\rho=9ah^2/8$ is the strength of the hydrodynamic interactions dictated by the functions
\begin{align}
S_{ij}&=\frac{4}{d_{ij}^3}\lb[\Theta_{ij}\cos\Delta_{ij}+\Phi_{ij} \cos\varphi_{ij}\rb],\\
L_{ij}&=\frac{4}{d_{ij}^3}\lb[\Theta_{ij}\sin\Delta_{ij}+\Phi_{ij} \sin\varphi_{ij}\rb].
\end{align}
Here we've defined phase difference $\Delta_{ij}=\phi_i-\phi_j$ and sum $\varphi_{ij}=\phi_i+\phi_j$ while the distance between cilium is given by $d_{ij}=2\sin(|\theta_{ij}|/2)$ where $\theta_{ij}$ is the angle between the cilia $i$ and $j$. Finally the functions
\begin{align}
\Theta_{ij}=\dfrac{\cos|\theta_{ij}|+\sin(|\theta_{ij}|/2)}{1+\sin(|\theta_{ij}|/2)},\\
\Phi_{ij}=\dfrac{\cos|\theta_{ij}|-\sin(|\theta_{ij}|/2)}{1+\sin(|\theta_{ij}|/2)},
\end{align}
capture the effect of the geometry of the spherical body on the hydrodynamic interactions, as shown in Fig.~\ref{Theta}.

We observe that as expected hydrodynamic interactions above a wall scale as ${O}(d_{ij}^{-3})$ \cite{vilfan2006,lenz2006}, but now, due to the spherical shape of the ciliate body, both the relative position, $\Delta$ and average position $\varphi$, of the two cilia around the ciliate body affect the hydrodynamic interactions as well by way of the geometric functions $\Theta$ and $\Phi$ respectively. The background flow velocity on each cilium (induced by the other) directly impacts both the angular velocity of each cilium, through the function $S_{ij}$, as well as its radial position via the function $L_{ij}$, which in turn affects the phase-speed as well.

\begin{figure}
\begin{center}
\includegraphics[scale=0.65]{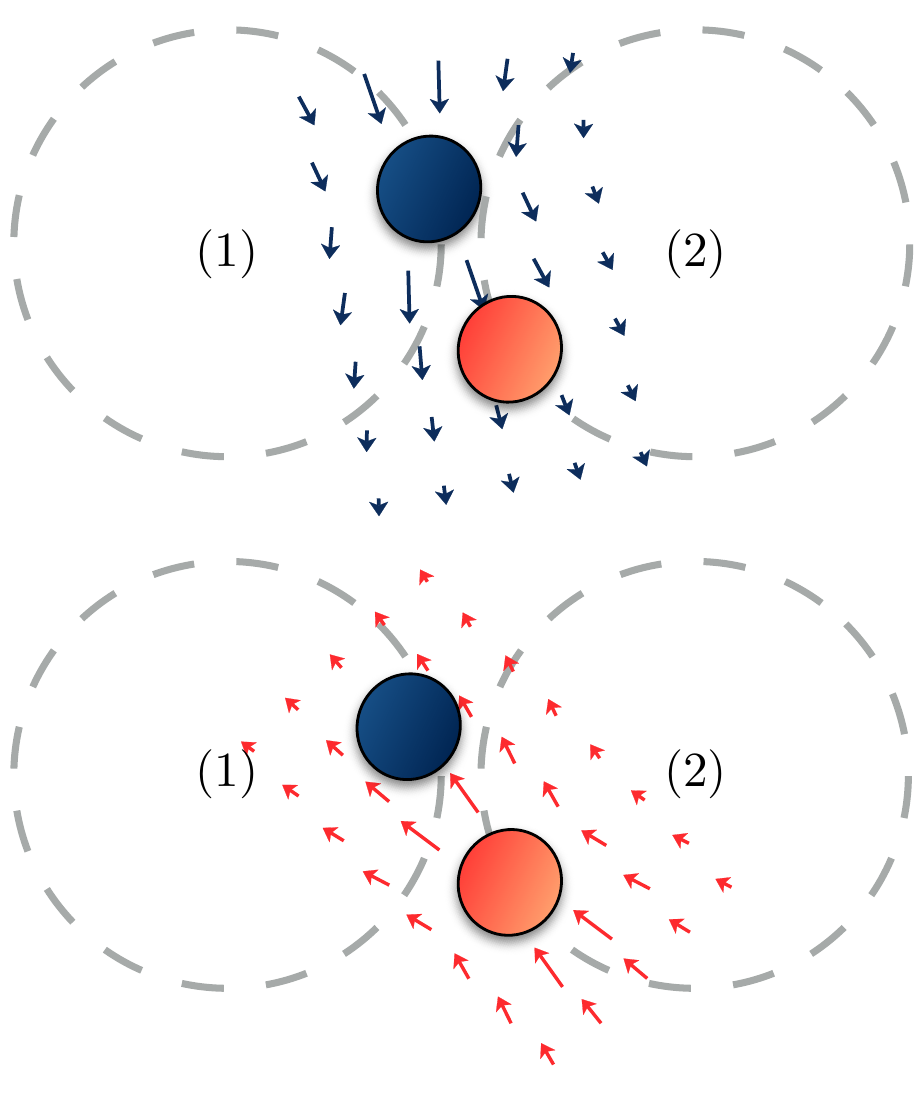}
\caption{The effect of the background flow field on the motion of each cilia. The two cilia are orbiting clockwise, cilium (2) is ahead, thus its induced flow field pulls cilium (1) to a smaller radius of trajectory which increases the instantaneous velocity of cilium (1). On the other hand, the velocity of cilium (2) decreases as the flow field of cilium (1) pushes cilium (2) to a larger trajectory. In this figure arrows show the flow field induced by each cilium.}
\label{locking}
\end{center}
\end{figure}

Taking the difference of \eqref{motion-1} and \eqref{motion-2} we obtain an evolution equation for the phase difference
\begin{align}
\label{non-ave}
\dot{{\Delta}}_{12}=\Delta\omega(1-\rho S_{12})-2(\rho/\kappa)\omega_1\omega_2{L}_{12},
\end{align}
where $\Delta\omega=\omega_1-\omega_2$. We see the phase difference evolves due to a difference in the intrinsic phase-speed, due to hydrodynamic interactions directly but also indirectly because of elastic radial displacements. To illustrate the latter point, let us assume cilium (2) is ahead by a positive phase difference of $\Delta_{12}$, as shown in Fig.~\ref{locking}. In this case, the background flow drives a contraction of the orbit for cilium (1) ($R_1<R_0$) and expansion of the orbit for cilium (2) ($R_2>R_0$). Since the internal driving forces of the cilia are constant, the changes in trajectories speed up cilium (1) and slow down cilium (2).

If intrinsic velocities are different, $\Delta\omega\ne 0$, for an equilibrium phase difference to arise, this difference must not overwhelm the elasto-hydrodynamic interactions, in other words $\Delta\omega= O(\rho)$ for fixed points in phase-difference. The second term on the right-hand side of equation \eqref{non-ave} is then $O(\rho^2)$ and shall be neglected.

To provide further clarity we note that the individual phases evolve on a much shorter time scale than the phase-differences, $\dot{\phi}_i= O(1)$ while $\dot\Delta_{12}=O(\rho)$; hence, we use a multiple scale analysis and average over a period of the short-time scale, $\tau_{_{\phi}}=2\pi/\omega_1 +O(\rho)$ to focus on the long-time behavior of the phase difference (indicated with an overbar). The cycle-averaged evolution equation for the phase-difference is then an Adler equation
\begin{align}
\label{two-sync}
\dot{\bar{\Delta}}_{12}=\Delta\omega-\gamma\Theta_{12}\sin\bar{\Delta}_{12},
\end{align}
where the synchronization strength in the case of a flat wall $\gamma = 8(\rho/d^3\kappa)\omega_1\omega_2$ \cite{niedermayer2008}, is augmented by the geometric term $\Theta_{12}$. If the frequency mismatch is small enough to be balanced by the elasto-hydrodynamic coupling, $\lb|\Delta\omega\rb|<\gamma\Theta_{12}$, a steady-state phase difference emerges given by $\bar{\Delta}^\eq_{12}=\sin^{-1}(\Delta\omega /\gamma\Theta_{12})$. In the limiting case of a rigid cilium ($\kappa\rightarrow \infty$) hydrodynamic interactions do not lead to an evolution of phase, and no phase-locking can occur where $\Delta\omega\ne 0$. When the cilia are identical (i.e., $\omega_1=\omega_2=1$) the phase-locking of the system is guaranteed (if $\theta_{12}\neq\pi$) as equation \eqref{two-sync} reduces to
 \begin{align}
 \label{two-identical}
\dot{\bar{\Delta}}_{12}=-\gamma\Theta_{12}\sin\bar{\Delta}_{12},
 \end{align}
indicating that the equilibrium phase difference is zero.

Unsurprisingly, the evolution equations for phase difference on a spherical body are largely similar to above a flat wall under the assumption that the cilia are much smaller than the ciliate. The difference is that the hydrodynamic interactions are mediated by the geometry of the body through $\Theta_{12}$. We see that when the two cilia are located at the opposite sides of the spherical body ($\theta_{12}=\pi$), radial interactions are completely screened by the ciliate as $\Theta_{12}=0$. We also note that for the angles near zero (and $2\pi$), special care must be used as these limits force $d_{12}\rightarrow 0$. To evaluate the system at these angles we can rescale the evolution equations \eqref{motion-1} and \eqref{motion-2} with distance $\hat{d}_{12}$, thereby recovering the flat body solution reported in Ref. \cite{niedermayer2008} in the limiting case where $\theta_{12}\rightarrow 0~(\text{or}~2\pi)$ and $\hat{A}\rightarrow\infty$.

\section{Interactions of chain of cilia}
Now we proceed to the system of $N$ identical cilia around a spherical body where $N\geq 3$. Relying on the linearity of the Stokes equation, the flow field induced by a chain of cilia can be determined by summing the contributions of all the cilia as well as their image points. Following the procedure outlined in the case of two cilia, the evolution equation for cilium (i) in a chain of $N$ cilia, to the leading order, is 
\begin{align}
\label{chain}
\dot{\phi}_{i}&= \omega_{i}+ \rho\sum_{j\neq i}^N\omega_{j}{S}_{ij} - \dfrac{\rho\omega_{i}}{\kappa} \sum^N_{j\neq i}\omega_{j}{L}_{ij},
\end{align}
where $\{i,j\}\in\{1,2, \ldots, N\}$. For simplicity, we shall assume first that all cilia in the chain have the same intrinsic angular velocity, $\omega_{i}=\omega_j=1$. The evolution of the phase difference on the long time scale is then
\begin{align}
\label{chain2}
{\dot{\bar{\Delta}}}_{1i}&=4 \rho\sum_{m\neq 1}^N \sum_{j\neq i}^N \lb(d_{1m}^{-3}\Theta_{1m}\cos\Deltab_{1m}-d_{ij}^{-3}\Theta_{ij}\cos\Deltab_{ij}\rb) \nonumber\\
&- \dfrac{4\rho}{\kappa} \sum_{m\neq 1}^N\sum^N_{j\neq i}\lb(d_{1m}^{-3}\Theta_{1m}\sin\Deltab_{1m}-d_{ij}^{-3}\Theta_{ij}\sin\Deltab_{ij}\rb),
\end{align}
where we have set cilium (1) as the reference phase, defining ${{{\Delta}}}_{1i}=\phi_1-\phi_{i}$. We note that unlike the case of two identical cilia where the average tangential terms do not contribute to synchronization (because of a pair-wise symmetry), in the case of many cilia the tangential terms ($S_{ij}$) do not vanish and hence contribute to evolution of the phase differences.

To further simplify the system, we now consider a chain of cilia which are equally distributed around the body and hence the angle between any two cilia is $\theta_{ij}=2\pi(i-j)/N$. By direct substitution into equation \eqref{chain2}, one can show that an equal phase-difference, $\Delta^\eq$, between all neighboring cilia is a fixed point of equation \eqref{chain2}. 
Because the system is periodic (in $\theta$), the sum of the phase differences must be an integer multiple of $2\pi$, 
\begin{align}
\label{fixed-point-2}
\Delta^\eq N=2\pi M,
\end{align}
where $M\in \mathbb{Z}$ for which there are only $N$ unique solutions (due to periodicity in $\phi$), synchronized (when $M=0$) or metachronal waves (when $M\neq0$).

\begin{figure}
\begin{center}
\includegraphics[scale=0.3]{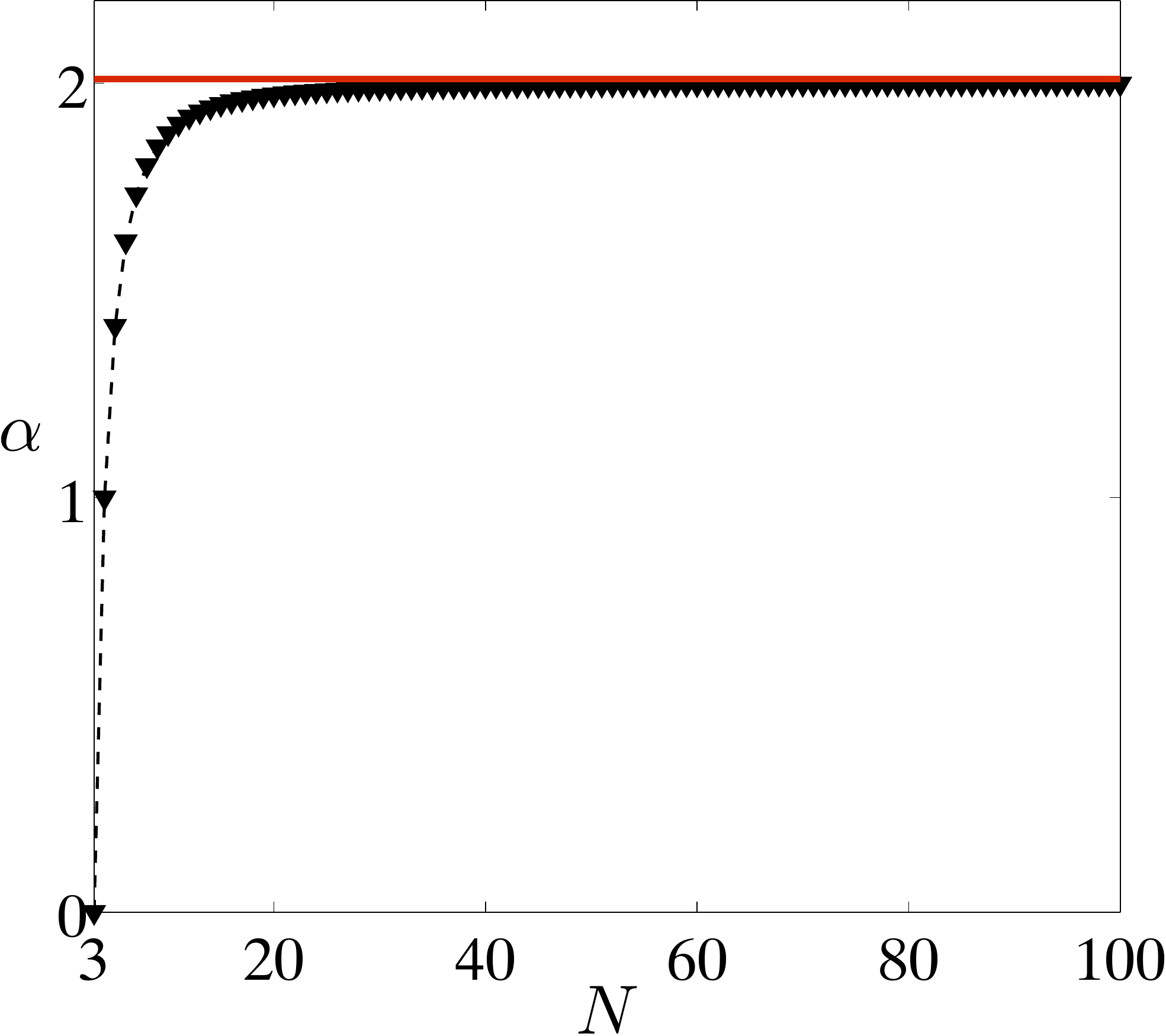}
\caption{The value of the coefficient $\alpha$, which dictates the stability of fixed points of a ciliary chain, is shown as a function of the given number of cilia $N$. }
\label{alpha}
\end{center}\end{figure}

The strength of the interactions between a pair of cilia decays rapidly as their distance increases, due both to the $d_{ij}^{-3}$ term as well as the effect of the angle, thus, we perform a linear stability analysis of these equilibrium states of the system considering only nearest neighbor interactions. Without loss of generality, we assume $-\pi\leq\Delta^\eq\leq\pi$. Using Gaussian elimination, we determined the maximum eigenvalues of the Jacobian at $\Delta^\eq$, as
\begin{align}
 \lambda_1=4d_{12}^{-3}\Theta_{12}\Big[(\rho/\kappa)(& \alpha\text{sgn}[\cos\Delta^\eq]-2)\cos\Delta^\eq
+\alpha\rho\sin|\Delta^\eq|\Big],
 \label{max}
 \end{align}
where $\alpha(N)\in[0,2)$ is a constant which depends on the number of cilia as shown in Fig.~\ref{alpha}. As an example, when $N=3$ the angle between each pair of cilia is $2\pi/3$. In this case $\alpha=0$ hence the system has a stable equilibrium only if $\cos\Delta^\eq>0$ and of the possible solutions $\Delta^\eq = {0,\pm 2\pi/3}$ only $\Delta^\eq=0$ is stable. When the cilia are all in-phase, $\Delta^\eq=0$, $\lambda_1<0$ for any $N$, indicating asymptotic stability of the synchronized state for any number of identical, evenly distributed cilia on a spherical ciliate provided the system has finite flexiblity. In the rigid limit, $\kappa\rightarrow\infty$, the largest eigenvalue is zero which causes a loss of stability of the synchronized state (as shown in numerical simulations).

\begin{figure}
\begin{center}
\includegraphics[scale=0.49]{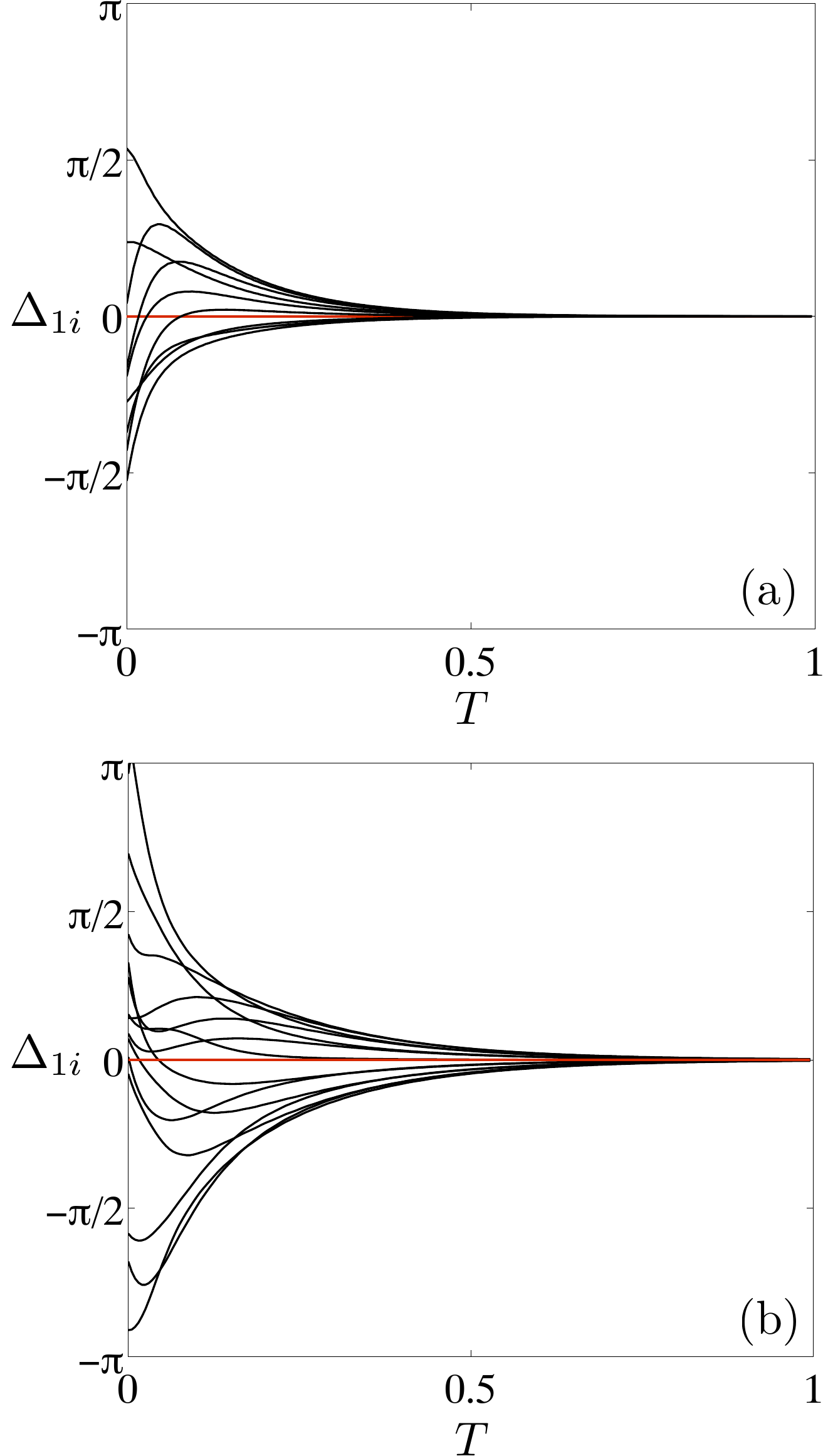}
\caption{Synchronization of a chain of (a) 10 and (b) 15 identical cilia distributed uniformly around a spherical body, with the random initial phases. Each line indicates the evolution of the phase difference for each cilium $i$ compared to cilium (1), $\Delta_{1i}=\phi_1-\phi_i$, over the time $T=t/(\kappa/\rho)$. These plots are the numerical evaluation of equation \eqref{chain} at the characteristic values of $\rho=3.6\times10^{-6}$, $\kappa=100$ and $\bar{\hat{\omega}}=20\pi~\text{rad}.s^{-1}$ \cite{okada1999,camalet2000}.}
\label{sync-sync}
\end{center}
\end{figure}

For a system to form metachronal waves, a non-zero equilibrium phase difference between the cilia is required. However, one can show directly that if $\pi/2\leq|\Delta^\eq|\leq\pi$, $\lambda_1>0$ while when $0\leq|\Delta^\eq|<\pi/2$ for stability one must have the integer
\begin{align}
M<\dfrac{N}{2\pi}\tan^{-1}\lb(\dfrac{2-\alpha}{\alpha\kappa}\rb),
\end{align}
which is satisfied only by $M=0$ for $\kappa>1$ (and here $\kappa \gg 1$). Thus, all non-zero values of $M$ (metachronal waves) are linearly unstable for any number of identical, evenly distributed cilia on a spherical ciliate. Unlike the reported results for the chain of cilia near a flat boundary \cite{niedermayer2008}, this system cannot form a stable metachronal wave and all the cilia eventually synchronize. The synchronization of two chains of $N=10$ and $N=15$ cilia from random initial conditions has been illustrated numerically in Fig.~\ref{sync-sync}.

\begin{figure}
\begin{center}
\includegraphics[width=0.37\textwidth]{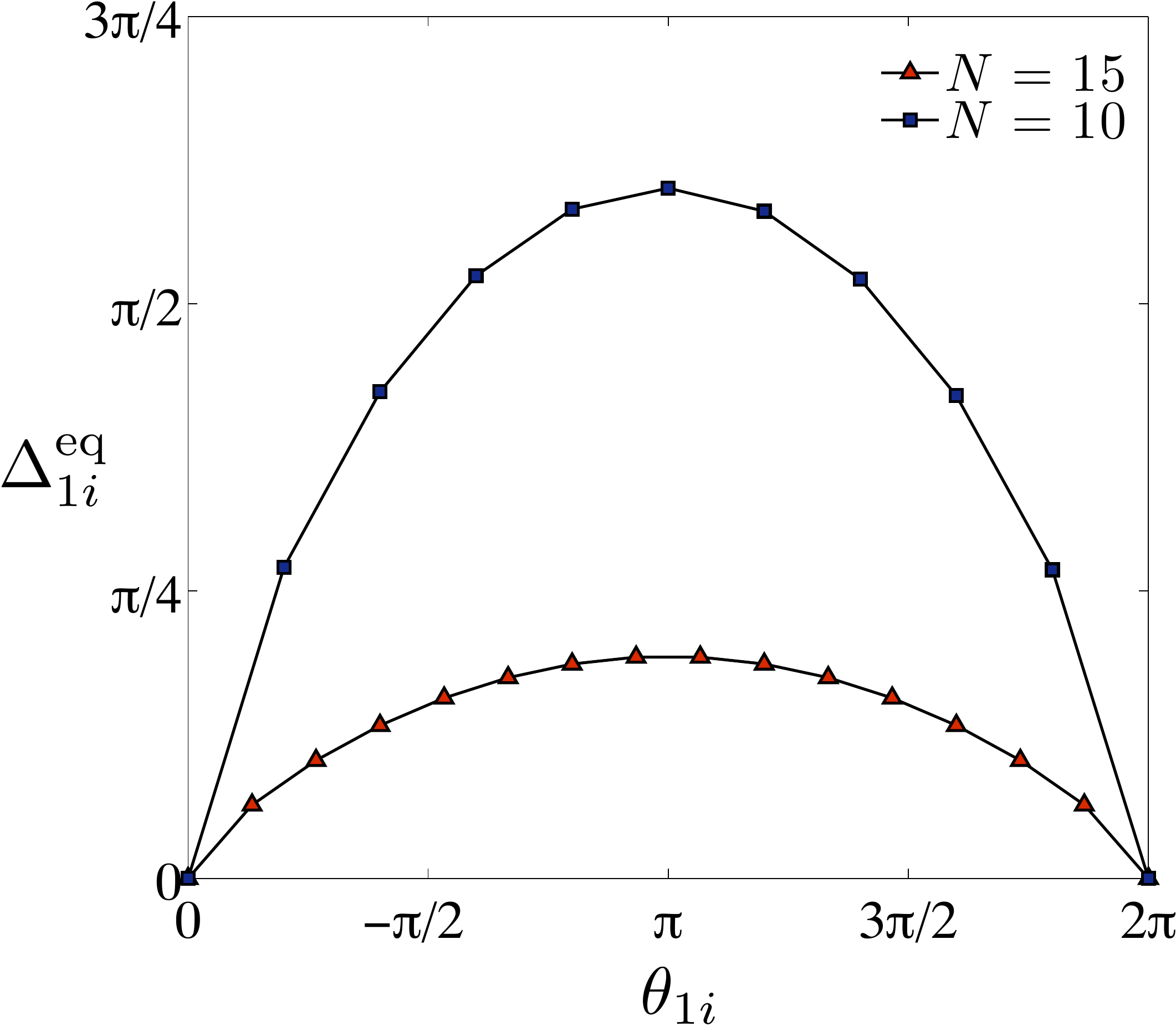}
\caption{Phase differences of nearby cilia in a chain of 10 and 15 cilia around a spherical body when the intrinsic angular velocity of cilium (1) is higher compared to the other cilia by $\Delta\omega=10^{-6}$ for both cases.}
\label{wave}
\end{center}
\end{figure}

In real biological examples one can hardly expect perfect symmetry and uniformity in the system so it is important to understand the effect of an imposed asymmetry on the stability of this system. There are several well documented sources of asymmetry, from biochemical noise \cite{goldstein2009,wan2014}, to the different intrinsic properties of a developing cilium \cite{hagiwara2004,goldstein2011} or even the addition of a transverse external flow \cite{guirao2007} which have all been found to spontaneously affect the behavior of a ciliary system. In particular, the beating rate of a developing cilium fluctuates as it grows, which can perturb the equilibrium state of the system \cite{hagiwara2004}. To analyze this phenomenon, we impose an asymmetry on the system by increasing the intrinsic velocity of cilium (1) to $1+\Delta\omega$, where we assume $\Delta\omega\ll 1$. The evolution equations for the phase differences are then
\begin{align}
\label{chain3}
{\dot{\bar{\Delta}}}_{1i}=\Delta\omega
&+4 \rho\sum_{m\neq 1}^N \sum_{j\neq i}^N \lb(d_{1m}^{-3}\Theta_{1m}\cos\Deltab_{1m}-d_{ij}^{-3}\Theta_{ij}\cos\Deltab_{ij}\rb) \nonumber\\
&-\dfrac{4\rho}{\kappa} \sum_{m\neq 1}^N\sum^N_{j\neq i}\lb(d_{1m}^{-3}\Theta_{1m}\sin\Deltab_{1m}-d_{ij}^{-3}\Theta_{ij}\sin\Deltab_{ij}\rb).
\end{align}
Now, due to the imposed asymmetry, the system no longer converges to a synchronized state where phases are equal. 
There must be a non-zero equilibrium phase difference between cilium (2) and cilium (1) (for example) to balance the difference in the intrinsic velocities. However, the effect of the imposed asymmetry becomes weaker as the distance from cilium (1) increases and, therefore, the phase difference between a two adjacent cilia decreases. These phase differences form a coordinated system of beating, which is illustrated in Fig.~\ref{wave} for two sample cases of $N=10$ and $N=15$.

These results indicate that the system responds to this asymmetry through an emergent wave-like behavior. Since this asymmetry arises from any developing cilium in the chain, these waves can originate from different parts of the ciliate and vanish once the beating rate of the developing cilium reaches the frequency of the other cilia. Here we should note that unlike metachronal waves which have $\sim 7-10$ cilia per wavelength \cite{machemer1972,wollin2011}, the asymmetry-induced behavior has a characteristic wavelength which spans the entire chain of $N$ cilia. Furthermore, as $N$ increases, the strength of the imposed asymmetry becomes weaker and the equilibrium phase differences of the cilia decrease. Thus, the amplitude of such waves is inversely proportional to the number of cilia in the chain, as shown in Fig.~\ref{wave}.

\section{Conclusion}
In this paper, we used a minimal model to capture the dynamics of cilia on a spherically shaped microorganism. We showed that, similar to the case of cilia above a flat wall, elasto-hydrodynamic interactions can lead to synchronization, however here the interactions are additionally mediated due to the geometry of the ciliate body.

For a chain of identical cilia uniformly distributed around a spherical boundary, we showed that the natural periodicity in the geometry of the ciliate leads the system to synchronize. We also showed that in this system, metachronal waves are strictly unstable fixed points of the dynamical system unlike in the case of interaction above a flat wall. This result suggests that the geometry of the ciliate plays a crucial role in the behavior of the ciliary chain and it has to be accounted for when analyzing microorganisms with curved bodies and suggests that a natural extension of this analysis would be to investigate a distrubtion of cilia over the whole surface of the ciliate. Our results also suggest that to form stable metachronal waves, rotation and translation of the ciliate \cite{friedrich2012}, elasticity of the cell-internal fibers connecting the cilia \cite{quaranta2015}, or motion of the cilia perpendicular to the ciliate body \cite{brumley2012}, may be necessary in such microorganisms. We also reported a wave-like response of the system when one of the cilia is intrinsically faster. In this case, the neighboring cilia display stable phase-locking with the faster cilium with a phase difference that decreases with distance from the asymmetry. Although the characteristics of this asymmetry-induced phenomenon do not match metachronal waves, we should note that in real ciliary chains there are likely many cilia of differing lengths or biochemical noise which may lead to more complex dynamics in biological systems.

\begin{acknowledgements}
The authors thank Professor G.M. Homsy for helpful discussions and support. G.J.E. acknowledges funding from NSERC.
\end{acknowledgements}

\bibliography{reference}

\end{document}